\documentclass[12pt,a4]{article}

\usepackage[dvipdfm]{graphicx}
\usepackage[dvipdfm]{color}
\usepackage[dvipdfm]{hyperref}
\usepackage{oldgerm}
\usepackage{yfonts}
\usepackage{euler}
\usepackage{amsmath}
\usepackage{amssymb}
\usepackage{amstext}
\usepackage{amscd}
\usepackage{amsfonts}

\begin{document}

\title{{\bf A consequence of the repulsive Casimir-Lifshitz force on nano-scale and the related Wheeler propagator in the classical electrodynamics}}
\author{Ferenc M\'arkus\thanks{e-mail: markus@phy.bme.hu, markusferi@tvn.hu
(corresponding author)}, Mario Rocca\thanks{email: rocca@fisica.unlp.edu.ar} and Katalin 
Gamb\'ar\thanks{e-mail: gambar@gdf.hu, gakati@gmail.com} \\ \\
$^{*}$Deparment of Physics, Budapest University of Technology
\\ and Economics, \\ Budafoki \'ut 8., H-1521 Budapest, Hungary \\
$^{\dag}$Departamento de F\'{\i}sica, Faculdad de Ciencias Exactas, \\
Universidad Nacional de La Plata, C.C. 67, 1900 La Plata, Argentina \\
$^{\ddag}$Institute of Basic and Technological Sciences, \\ Dennis Gabor
College, \\ M\'ern\"ok u. 39., H-1119 Budapest, Hungary}

\maketitle

\begin{abstract}
Mainly on nano-scale, but maybe not exclusively, it can be imagined a spontaneous charge 
disjunction inside certain media due to the fluctuations, collisions, wall effects, radiation 
or/and other presently unknown interactions. The repulsive Casimir-Lifshitz force may be
also a good candidate as a source of this kind of phenomenon. To cover this assumption
mathematically an additional term should appear in the Maxwell equations. As a consequence of 
this term, similar Klein-Gordon equations with a negative mass term will be obtained for the
electric and magnetic fields. These equations may have tachyonic solutions depending on the 
parameter of the charge disjunction process. Finally, the propagator is formulated, the electric
field and the charge density are calculated during the charge disjunction.
\end{abstract}

\fontsize{12}{18}\selectfont \renewcommand{\thesection}{\arabic{section}}

\section{Introduction}

In the present paper the whole description remains within the framework of the classical 
electrodynamics. We assume that there are no net external  currents and charges, i.e., if there 
are currents and charges these appear as a consequence of the internal processes inside the 
electric conductive medium. Assumable, the source of the mentioned processes maybe fluctuations, 
collisions, wall interactions, thermal effects, radiation. Presently, just as an interesting idea 
we may imagine this charge disjunction due to a repulsive Casimir-Lifshitz(-like) force 
\cite{munday,lamoreaux}. If we accept this possibility we should add a term in the Maxwell
equations to establish the relevant mathematical equations. The calculation shows that a 
Klein-Gordon equation with a negative mass term will be obtained for both the electric and the 
magnetic field. There have been pointed out in a previous paper \cite{mg2010} that applying 
Feynman's and Wheeler's ideas to create the so-called Wheeler propagator, and the mathematical 
tools and steps ingeniously developed by Bollini, Rocca and Giambiagi \cite{bollini99,bollini98,
bollini96,bollini04,bollini97a,bollini97b} based on the Bochner's theorem \cite{bochner,jerri} we 
can find the propagator for a similar field equation, namely, for the Lorentz invariant thermal 
energy propagation \cite{mg2010,gm2007}. Since, the present mathematical equations for the electric 
and magnetic fields apart from the parameters are completely the same thus the mathematical 
elaboration is also the same. Consequently, we can obtain the causal propagator of the 
electromagnetic process and finally we calculate the electric field to demonstrate how the charge 
density increases in time.

\section{Repulsive Casimir-Lifshitz force in the electrodynamics}

In the present calculations we can assume that a repulsive Casimir-Lifshitz(-like) force 
\cite{munday,lamoreaux} may cause this kind of charge motion. The spontaneous polarization process 
starts at $t_{0}$ and ends $t$, and the time ${\tau}=t-t_{0}$ is really very short. The aim is to 
formulate those kind of equations of motion that preserve the Lorentz invariance of the theory and 
finally the description shows the charge disjunction. We start from the Maxwell equations modifying
the second one (Eq. (\ref{Maxwell_2})) with an additional term.

\begin{subequations}
\begin{equation}  \label{Maxwell_1}
\frac{1}{{\mu}_{0}} curl {\bf B} = {\varepsilon}_{0} \frac{\partial{\bf E}}{{\partial}t} + {\bf J},
\end{equation}
\begin{equation}  \label{Maxwell_2}
curl {\bf E} = - \frac{\partial{\bf B}}{{\partial}t} + {\alpha}^{2} \int_{t_0}^{t} {\bf B}({\bf r},t')dt',
\end{equation}
\begin{equation}  \label{Maxwell_3}
{\varepsilon}_{0}div {\bf E} = {\varrho},
\end{equation}
\begin{equation}  \label{Maxwell_4}
div {\bf B} = 0.
\end{equation}
\end{subequations}

\noindent At this point, maybe, this step is not obvious, but it will be shown that the third term in Eq.
(\ref{Maxwell_2}) generates the charge disjunction. This process can be considered as the consequence
of an internal repulsive interaction similarly to the repulsive Casimir-Lifshitz force 
\cite{munday,lamoreaux}. In order to solve these equations the vector potential ${\bf A}$ is 
introduced by the help of Eq. (\ref{Maxwell_4}) as

\begin{equation}
{\bf B} = curl{\bf A}.
\end{equation}

\noindent Substituting this in to Eq. (\ref{Maxwell_2}) and rearranging the obtained formula we get

\begin{equation}
curl \left( {\bf E} - \frac{\partial{\bf A}}{{\partial}t} - {\alpha}^{2} \int_{t_0}^{t}
{\bf A}({\bf r},t')dt' \right) = 0.
\end{equation}

\noindent We can express the electric field ${\bf E}$ from the above equation

\begin{equation}  \label{field_E}
{\bf E} = -\frac{\partial{\bf A}}{{\partial}t} - grad \, {\varphi}  + {\alpha}^{2} \int_{t_0}^{t}
{\bf A}({\bf r},t')dt',
\end{equation}

\noindent where we introduces the scalar potential $\varphi$. We note that this formula is not
Lorentz invariant, however, this fact will not cause serious problem in the description since 
the primary field variables are the vector and scalar potentials $A_{\mu}=({\varphi},{\bf A})$.
Now, we take Eq. (\ref{Maxwell_3}) and we replace ${\bf E}$ into it, thus we can write

\begin{equation}  \label{div_E}
div{\bf E} = - \frac{{\partial}(div{\bf A})}{{\partial}t} - {\Delta}\varphi +
{\alpha}^{2} \int_{t_0}^{t} div{\bf A}({\bf r},t')dt' = \frac{\varrho}{{\varepsilon}_{0}}.
\end{equation}

\noindent We apply the Lorentz gauge taking ${\varepsilon}_{0}{{\mu}_{0}}=1$

\begin{equation}  \label{lorentz_gauge}
\frac{\partial\varphi}{{\partial}t} + div{\bf A} = 0,
\end{equation}

\noindent we can eliminate the vector potential in Eq. (\ref{div_E}), thus we obtain

\begin{equation}  \label{K-G_eq_varphi}
\frac{{\partial}^{2}\varphi}{{\partial}t^{2}} - {\Delta}\varphi - {\alpha}^{2}\varphi =
\frac{\varrho}{{\varepsilon}_{0}}.
\end{equation}

\noindent This is a Lorentz invariant Klein-Gordon equation with negative mass term. Its structure 
is similar to the equation in Refs. \cite{mg2010,gm2007} that describes a dynamical phase 
transition as a consequence of a spinodal instability. The construction of Wheeler propagator is 
originated from the similar group of phenomena, the interaction of the radiation and the absorbing 
media \cite{bollini99,feynman1,feynman2}. \\
\noindent The equation for the vector potential can be also formulated, starting from Eq.
(\ref{Maxwell_1}) and substituting the form of electric field from Eq. (\ref{field_E})

\begin{equation}
curl \, curl{\bf A} = - \frac{{\partial}^{2}{\bf A}}{{\partial}t^{2}} -
\frac{{\partial}grad \varphi}{{\partial}t} + {\alpha}^{2}{\bf A} + {\mu}_{0}{\bf J}.
\end{equation}

\noindent Applying the vector identity $curl \, curl = grad \, div - \Delta$ and the Lorentz gauge
in Eq. (\ref{lorentz_gauge}) we can rewrite the above equation in a more expressive form

\begin{equation}  \label{K-G_eq_A}
\frac{{\partial}^{2}{\bf A}}{{\partial}t^{2}} - {\Delta}{\bf A} - {\alpha}^{2}{\bf A} = 
{\mu}_{0}{\bf J},
\end{equation}

\noindent which is also a Lorentz invariant Klein-Gordon equation with a negative mass term for 
the vector potential. Both the scalar and the vector potentials as basic fields -- the components of 
a four vector $A_{\mu}=({\varphi},{\bf A})$ -- fulfill Lorentz invariant equations with the same
structure, they propagates with the same speed, thus the whole description is Lorentz invariant. \\

\noindent Now, we should write the equations for the field variables ${\bf E}$ and ${\bf B}$. Thus,
we take the time derivative of Eq. (\ref{Maxwell_1})

\begin{equation}
curl \frac{{\partial}\bf B}{{\partial}t} = \frac{{\partial}^{2}{\bf E}}{{\partial}t^{2}} +
{\mu}_{0} \frac{{\partial}{\bf J}}{{\partial}t}.
\end{equation}

\noindent The term of the left hand side can be substituted after the rotation of Eq. 
(\ref{Maxwell_2}) by which we write

\begin{equation}
curl \left(-curl{\bf E} + {\alpha}^{2} \int_{t_0}^{t} {\bf B}({\bf r},t')dt' \right) =
\frac{{\partial}^{2}{\bf E}}{{\partial}t^{2}} + {\mu}_{0} \frac{{\partial}{\bf J}}{{\partial}t}.
\end{equation}

\noindent We can eliminate the field ${\bf B}$ applying again Eq. (\ref{Maxwell_1}), and finally
we obtain

\begin{equation}  \label{K-G_eq_E}
\frac{{\partial}^{2}{\bf E}}{{\partial}t^{2}} - \Delta{\bf E} - {\alpha}^{2}{\bf E} 
= -\frac{1}{{\varepsilon}_{0}} grad \, {\varrho} - {\mu}_{0} \frac{{\partial}{\bf J}}{{\partial}t}
+ {\mu}_{0}{\alpha}^{2} \int_{t_0}^{t} {\bf J}({\bf r},t')dt'.
\end{equation}

\noindent Similarly, for the field ${\bf B}$, we take the time derivative of Eq. {\ref{Maxwell_2})

\begin{equation}
curl \frac{{\partial}{\bf E}}{{\partial}t} = - \frac{{\partial}^{2}{\bf B}}{{\partial}t^{2}} + 
{\alpha}^{2} {\bf B},
\end{equation}

\noindent and eliminating the field ${\bf E}$ by the help of the rotation of Eq. (\ref{Maxwell_1})
we get

\begin{equation}  \label{K-G_eq_B}
\frac{{\partial}^{2}{\bf B}}{{\partial}t^{2}} - \Delta{\bf B} - {\alpha}^{2}{\bf B}
= {\mu}_{0} curl \, {\bf J}.
\end{equation}

\noindent It can be seen that for all of the field equations, Eqs. (\ref{K-G_eq_varphi}), 
(\ref{K-G_eq_A}), (\ref{K-G_eq_E}) and (\ref{K-G_eq_B}), have the same structure. We know from
former studies \cite{gm2007,gm2008} that these Klein-Gordon equations with a negative mass term 
are resulted from repulsive interactions. Thus it seems to us that the interaction in the present
case is a repulsive Casimir-Lifshitz(-like) force which appears mathematically in the second
Maxwell-equation, in Eq. (\ref{Maxwell_2}).

\section{The Wheeler propagator and the time-evolution of the electric field}

Now, we can examine the appearing electrical field due to the charge disjunction given by Eq. 
(\ref{K-G_eq_E}). Since the last two terms of this equation make rather complicated the solution
and assuming that the contribution of the current and the time derivative of the current can be 
negligible somehow, mainly in the initial time, we cut the right hand side to simplify the problem 
to

\begin{equation}  \label{cut_K-G_eq_E}
\frac{{\partial}^{2}{\bf E}}{{\partial}t^{2}} - \Delta{\bf E} - {\alpha}^{2}{\bf E} 
= -\frac{1}{{\varepsilon}_{0}} grad \, {\varrho}.
\end{equation}

\noindent We solve this equation applying the Green function method, thus the electric field
${\bf E}$ can be expressed

\begin{equation}
{\bf E}({\bf r},t) = \int -\frac{1}{{\varepsilon}_{0}} grad {\varrho}(x') \left[
\frac{1}{(2{\pi})^{4}} \int d^{4}k \frac{e^{ik(x-x')}}{k^{2} - {\alpha}^{2}}
\right] dx',
\end{equation}

\noindent where the expression

\begin{equation}  \label{Green_function}
G(x,x') = \frac{1}{(2{\pi})^{4}} \int d^{4}k \frac{e^{ik(x-x')}}{k^{2} -
{\alpha}^{2}}
\end{equation}

\noindent in the $\left[ ... \right]$ bracket is the Green function. (Here, the coordinate 
$x=(r,t)$ involves both the space and time coordinates.) The physical description of the propagator 
is based on Feynman's and Wheeler's original idea \cite{feynman1,feynman2}, the mathematical method 
to elaborate the calculations applying the Bochner's theorem \cite{bochner,jerri} is developed by 
Bollini, Rocca and Giambiagi \cite{bollini99,bollini98,bollini96,bollini04,bollini97a,bollini97b}. 
Considering these preliminaries the solution of Eq. (\ref{Green_function}) in the present case is 
completely similar to the solution of Eq. (15) in Ref. \cite{mg2010} with the obvious difference 
that there we need to write $\alpha$ instead of $m$, and the connection 
${\alpha} = \sqrt{{\mu}_{0}S_{p}}$ must be applied \cite{wheeler_propagator,gradshteyn} to obtain 
the Wheeler propagator $(n=4)$

\begin{equation}  \label{Wheelerfv}
W^{(4)}(x) = \frac{{\alpha}}{8\pi}
(x_{0}^{2}-r^{2})_{+}^{-\frac{1}{2}}
I_{-1} \left( {\alpha} (x_{0}^{2}-r^{2})_{+}^{\frac{1}{2}} \right),
\end{equation}

\noindent by the notations

\[
x_{+}^{\beta}=x^{\beta} \;\;\; for \;\;\; x > 0,
\]
\[
x_{+}^{\beta}=0 \;\;\; for \;\;\; x < 0.
\]

\noindent The above propagator fulfill the requirement of causality. Finally, we can express the 
electric field ${\bf E}({\bf r},t)$ by the calculated propagator

\begin{equation}  \label{evol_field_E}
{\bf E}({\bf r},t) = \int -\frac{1}{{\varepsilon}_{0}} grad {\varrho}(x')
W^{(4)}(x-x') dx'
\end{equation}

\noindent as a four dimensional convolution. Probably, the spontaneous polarization cannot be 
detected directly but there may be several consequences of it. The demonstration of the phenomenon 
seems rather difficult because there is no freely propagating particle during the excitation.

\section{Evolution of a nearly flat Gaussian charge distribution}

On the basis of the previous calculations we expect that a small perturbation can grow up to a 
huge dipole pair within the short time $\tau$, and it disappears similarly fast. As an example we 
examine the time evolution of a Gaussian charge distribution

\begin{equation}  \label{charge_distribution}
{\varrho}(x) = {\varrho}_{0} e^{-ar^2} \delta (t-t_{0}).
\end{equation}

\noindent If the parameter $a \sim 0$, the charge distribution can be considered practically 
homogeneous $(e^{-ar^2} \sim 1)$. The time-dependent factor have a sharp peak if $\epsilon$
tends to zero. So, if we consider the charge gradient for small values of $a$ and $\epsilon$ we 
obtain

\begin{equation}  \label{charge_gradient}
grad \, \varrho(x) = -2{\varrho}_{0} a \delta (t) {\bf r} e^{-ar^2} \,
\sim \, -2{\varrho}_{0} a \delta (t-t_{0}) {\bf r}.
\end{equation}

\noindent Here, we apply the "early" form of the Wheeler propagator [see Eq. (32) in Ref.
\cite{mg2010}]

\begin{equation}  \label{wheeler_propagator}
W(x)=\frac {Sgn (x_{0})} {16\pi^3}\int\limits_{-\infty}^{\infty}
\frac {\sin(p^2-\alpha^2+i0)^{\frac {1} {2}}x_0} {(p^2-\alpha^2+i0)^{\frac {1} {2}}} 
e^{i{p}\cdot{x}} d^3p
\end{equation}

\noindent for the further calculations. Substituting the charge gradient in Eq. 
(\ref{charge_gradient}) and the Wheeler propagator in Eq. (\ref{wheeler_propagator}) into the 
expression of Eq. (\ref{evol_field_E}) then we obtain the time evolution of the electric field

\begin{equation}
{\bf E}(x)=\frac {2\varrho_{0} a} {16\pi^3\varepsilon_{0}}Sgn(t-t_{0})
\int\limits_{-\infty}^{\infty}
\frac {\sin(p^2-\alpha^2+i0)^{\frac {1} {2}}(t-t_{0})} {(p^2-\alpha^2+i0)^{\frac {1} {2}}}
e^{i{p}\cdot({x}-{x}')}{x}'\; d^3p \; d^3x'.
\end{equation}

\noindent After the evaluation of the integral ans simplifying the mathematical expression the 
electric field can be analytically expressed

\[
{\bf E}(x)=\frac{\varrho_{0} a}{\varepsilon_0}{\bf r} \; Sgn(t-t_{0})
\frac {\sin(-{\alpha}^2+i0)^{\frac {1} {2}}(t-t_{0})} {(-{\alpha}^2+i0)^{\frac {1} {2}}} =
\]
\begin{equation}  \label{electric_field}
\frac {\varrho_{0} a} {\varepsilon_{0} \alpha}{\bf r} \sinh(\alpha |t-t_{0}|)\,\sim\,
\frac {\varrho_{0} a} {\varepsilon_{0} \alpha}{\bf r} e^{-ar^2} \sinh(\alpha |t-t_{0}|)
\end{equation}

\noindent It can be read out easily from this exact result that the magnitude of the electric  
field ${\bf E}$ follows an exponential behavior. The source of the huge electric field is the
enormously growing charge distribution

\begin{equation}  \label{charge_density}
{\varrho}(r,t) = \frac {\varrho_{0}a}{\alpha} (3-2ar^{2}) e^{-ar^2} \sinh(\alpha |t-t_{0}|) \,
\sim \, 3 \frac{\varrho_{0}a}{\alpha} e^{-ar^2} \sinh(\alpha |t-t_{0}|),
\end{equation}

\noindent which can be obtained by the Maxwell equation $div {\bf E} = {\varrho}/{\varepsilon_0}$.
It is interesting to see that if we integrate this charge density for the whole space the result is
always zero for all positive values of the parameter $a > 0$

\begin{equation}
\int_{0}^{\infty} {\varrho}(r,t) dV =
4 {\pi} \int_{0}^{\infty} \frac {\varrho_{0} a} {\alpha} r^{2} (3-2ar^{2}) e^{-ar^2} 
\sinh(\alpha |t-t_0|) \; dr= 0,
\end{equation}

\noindent i.e., the conservation law of electric charge is completed, there is only internal 
movement of the charges -- charge disjunction. Applying the continuity relation

\begin{equation}
\frac{\partial\varrho}{{\partial}t} + div \, {\bf J} = 0,
\end{equation}

\noindent and considering Eqs. (\ref{Maxwell_3}) and (\ref{electric_field}) we obtain the
current ${\bf J}$

\begin{equation}  \label{current}
{\bf J} = -{\varrho}_{0} \, a \, {\bf r} e^{-ar^2} Sgn(t-t_{0})\cosh(\alpha |t-t_{0}|) \, \sim \,
-{\varrho}_{0} \, a \, {\bf r} \, Sgn(t-t_{0})\cosh(\alpha |t-t_{0}|).
\end{equation}

\noindent Here, we note that calculating the dropped part of Eq. (\ref{K-G_eq_E})

\begin{equation}
- {\mu}_{0} \frac{{\partial}{\bf J}}{{\partial}t}
+ {\mu}_{0}{\alpha}^{2} \int_{t_0}^{t} {\bf J}({\bf r},t')dt'
\end{equation}

\noindent with the above solution of the current in Eq. (\ref{current}), we obtain zero. (The 
initial condition for the current: at time $t_{0}$ the current ${\bf J}=0$.) Thus, we can say that 
the obtained solution for the electric field in Eq. (\ref{electric_field}) and the for the charge 
density in Eq. (\ref{charge_density}) from the cut Klein-Gordon equation in Eq. 
(\ref{cut_K-G_eq_E}) can be considered as exact results. \\

The above results clearly show that -- let us say -- the positive charges are moving towards the 
origin, and reversely, the negative charges are moving towards the radius $r = 1.5$. Of course, 
the sign of the charges may be opposite. Naturally, the process must be restricted to 
nano-distances. Furthermore, the process stops after the very short time $\tau$, and it turns 
back, so finally the system reaches its originally homogeneous charge distribution again. The 
present calculations does not involve the possibility of a charge oscillations, but this process 
may be also realistic. This examination and discussion are challenges for a future work. \\

\noindent The time evolution of the charge density is shown in Fig. 1. Since the physical situation 
is spherically symmetric it is enough to demonstrate the increase of the charge density as a 
function of the radius $r$ in different time moments.

\begin{figure}[h]
\centering
\includegraphics[width=8 cm, height=4 cm]{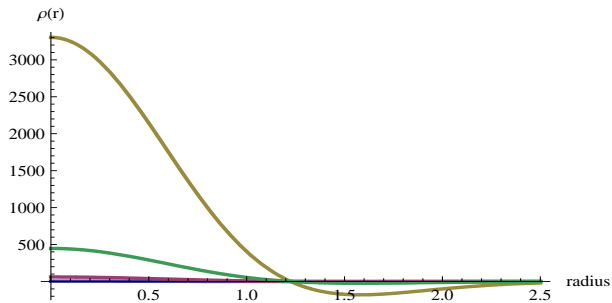}
\caption{Charge densities as a function of the radius $r$ in three different time moments: 
$t=0.3$ (purple line -- below), $t=0.6$ (green line -- middle), $t=0.8$ (grey line -- above). The 
charge density and time are in arbitrary units.}
\end{figure}

\noindent This figure demonstrates spectacularly how fast the charge density increases in time. The 
parameters can be taken optionally at the present stage, thus ${\varrho}_{0}=1, a=1$ and 
${\alpha}=10$, which means that the scale is arbitrary on the figure. We can see that at the 
beginning the charge density increases rather slowly comparing the later time, and in a certain 
time it can grow up in a giant form. During the elapsing time a negative spherically symmetric charge 
density is collecting with a maximal value at the radius $r=1.5$. The process stops at the very 
short time $\tau$, and it turns back, so finally the system reaches its originally homogeneous 
charge distribution. \\

\section{Summary}

In the present work it is pointed out that an assumed repulsive Casimir-Lifshitz-like force may
cause a giant charge disjunction on a short range within a short time. Inspite of the related
causal Wheeler propagator, probably, these excited particles are hidden non-observable entities.
However, if this process may happen, it may contribute to the physical behavior of some small 
systems, e.g. on nano-scale to the electric properties or other transport phenomena of few body 
systems.

\section{Acknowledgment}

The authors would like to thank the National Office of Research and Technology (NKTH; Hungary)
for financial support MX-20/2007 (Grant No. OMFB-00960/2008). This work is connected to the 
scientific program of the " Development of quality-oriented and harmonized R+D+I strategy and 
functional model at BME" project. This project is supported by the New Hungary Development Plan 
(Project ID: T?MOP-4.2.1/B-09/1/KMR-2010-0002).

\end{document}